\documentclass[english,aps,preprint,floatfix]{revtex4}
\usepackage[T1]{fontenc}
\usepackage[latin9]{inputenc}
\usepackage{array}
\usepackage{multirow}
\usepackage{graphicx}

\makeatletter

\providecommand{\tabularnewline}{\\}

\@ifundefined{textcolor}{}
{%
 \definecolor{BLACK}{gray}{0}
 \definecolor{WHITE}{gray}{1}
 \definecolor{RED}{rgb}{1,0,0}
 \definecolor{GREEN}{rgb}{0,1,0}
 \definecolor{BLUE}{rgb}{0,0,1}
 \definecolor{CYAN}{cmyk}{1,0,0,0}
 \definecolor{MAGENTA}{cmyk}{0,1,0,0}
 \definecolor{YELLOW}{cmyk}{0,0,1,0}
}

\makeatother

\usepackage{babel}
\begin{document}

\title{Production and decay of up-type and down-type new heavy quarks through
anomalous interactions at the LHC}

\author{\.{I}. Turk Çak\i r}

\email{ilkayturkcakir@aydin.edu.tr}

\selectlanguage{english}%

\affiliation{Istanbul Ayd\i n University, Application and Research Center for
Advanced Studies, 34295, Istanbul, Turkey }

\author{S. Kuday}

\email{sinankuday@aydin.edu.tr}

\selectlanguage{english}%

\affiliation{Istanbul Ayd\i n University, Application and Research Center for
Advanced Studies, 34295, Istanbul, Turkey }

\author{O. Çak\i r}

\email{ocakir@science.ankara.edu.tr}

\selectlanguage{english}%

\affiliation{Istanbul Ayd\i n University, Application and Research Center for
Advanced Studies, 34295, Istanbul, Turkey }

\affiliation{Ankara University, Department of Physics, 06100, Ankara, Turkey}
\begin{abstract}
We study the process $pp\to QVX$ (where $Q=t,b$ and $V=g,\gamma,Z$)
through the anomalous interactions of the new heavy quarks at the
LHC. Considering the present limits on the masses and mixings, the
signatures of the heavy quark anomalous interactions are discussed
and analysed at the LHC for the center of mass energy of $13$ TeV.
An important sensitivity to anomalous couplings $\kappa_{g}^{t'}/\Lambda=0.10$
TeV$^{-1}$, $\kappa_{\gamma}^{t'}/\Lambda=0.14$ TeV$^{-1}$, $\kappa_{Z}^{t'}/\Lambda=0.19$
TeV$^{-1}$ and $\kappa_{g}^{b'}/\Lambda=0.15$ TeV$^{-1}$, $\kappa_{Z}^{b'}/\Lambda=0.19$
TeV$^{-1}$, $\kappa_{\gamma}^{b'}/\Lambda=0.30$ TeV$^{-1}$ for
the mass of 750 GeV of the new heavy quarks $t'$ and $b'$ can be
reached for an integrated luminosity of $L_{int}=100$ fb$^{-1}$.
\end{abstract}
\maketitle

\section{introduction}

The standard model (SM) of the strong and electroweak interactions
describes succesfully the phenomena of particle physics. However,
there are many unanswered questions suggesting the SM to be an effective
theory. In order to answer some of the problems with the SM, additional
new fermions can be accommodated in many models beyond the SM (see
Refs. \cite{Holdom2009}, \cite{Atre2009}, \cite{Atre2011}, \cite{Chakdar2013}
and references therein). The new heavy quarks could also be produced
in pairs at the LHC with center of mass energy of $13$ TeV. However,
due to the expected smallness of the mixing between the new heavy
quarks and known quarks, the decay modes can be quite different from
the one relevant to charged weak interactions. A new symmetry beyond
the SM is expected to explain the smallness of these mixings. The
arguments given in Ref. \cite{Fritzsch1999} for anomalous interactions
of the top quark are more valid for the new heavy quarks $t'$ and
$b'$ due to their expected larger masses than the top quark. 

The ATLAS experiment \cite{ATLAS-3} and CMS experiment \cite{CMS-5}
have searched for the fourth generation of quarks and set limits on
the mass of $m_{t'}>570$ GeV and $m_{b'}>470$ GeV at $\sqrt{s}=7$
TeV. The pair production of new heavy quarks have been searched by
the ATLAS experiment \cite{Aad:2014efa}, \cite{ATLAS:2012qe} and
the $m_{t'}>656$ GeV mass limits are set at $\sqrt{s}=7$ TeV. The
CMS experiment have excluded $t'$ masses below 557 GeV\cite{CMS-6}
. The vector-like quarks have been searched by the ATLAS experiment
\cite{Aad:2011yn}, \cite{Aad:2012uu} and set bounds as $900$ GeV
for charged current channel and 760 GeV for neutral current channel
at $\sqrt{s}=7$ TeV. The CMS experiment \cite{CMS-7}, \cite{CMS-8}
have set the lower bounds on the mass of $687$ GeV at $\sqrt{s}=8$
TeV. Some of the final states in the searches of new phenomena \cite{ATLAS2014}
and excited quarks \cite{Aad:2013rna} can also be considered in relation
with the new heavy quarks. 

The anomalous resonant productions of the fourth family quarks have
been studied in Refs. \cite{Ciftci:2008tc,Cakir2009} at the LHC with
$\sqrt{s}=14$ TeV. sThe possible single productions of fourth generation
quarks via anomalous interactions at Tevatron have also been studied
in Refs. \cite{Arik:2002sg,Sahin}. The parameter space for the mixing
of the fourth generation quarks have been presented in Ref. \cite{Bobrowski}.
The CP violating flavour changing neutral current processes of the
fourth generation quarks have been analysed in Ref. \cite{Eilam2009},
and the large mixing between fourth generation and first three generations
have been excluded under the proposed fit conditions. Investigation
of the parameter space favoured by the precision electroweak data
have been performed for the fourth SM family fermions in Ref. \cite{OPUCEM}. 

In this work, we present the analysis of anomalous productions and
decays of new heavy quarks $t'$ and $b'$ at the LHC. We have performed
the fast simulation for the signal and background. Any observations
of the invariant mass peak in the range of $500-1000$ GeV and excess
in the events with the final states originating from $tV$ and $bV$
can be interpreted as the signal for the new heavy quarks $t'$ and
$b'$ via the anomalous interactions.

\section{Heavy Quarks Anomalous Interact\i ons}

A general theory that has the standard model (SM) as its low energy
limit can be written as a series in $\Lambda^{-1}$ with operators
obeying the required symmetries. The effective Lagrangian for the
anomalous interactions among the heavy quarks ($Q'\equiv t'$ or $b'$),
ordinary quarks $q$, and the gauge bosons $V=\gamma,Z,g$ can be
written explicitly:
\[
L=\sum_{q_{i}=u,c,t}\frac{\kappa_{\gamma}^{q_{i}}}{\Lambda}Q_{q_{i}}g_{e}\overline{t}'\sigma_{\mu\nu}q_{i}F^{\mu\nu}+\sum_{q_{i}=u,c,t}\frac{\kappa_{z}^{q_{i}}}{2\Lambda}g_{z}\overline{t}'\sigma_{\mu\nu}q_{i}Z^{\mu\nu}+\sum_{q_{i}=u,c,t}\frac{\kappa_{g}^{q_{i}}}{2\Lambda}g_{s}\overline{t}'\sigma_{\mu\nu}\lambda_{a}q_{i}G_{a}^{\mu\nu}
\]

\begin{equation}
\mbox{}\:\mbox{}+\sum_{q_{i}=d,s,b}\frac{\kappa_{\gamma}^{q_{i}}}{\Lambda}Q_{q_{i}}g_{e}\overline{b}'\sigma_{\mu\nu}q_{i}F^{\mu\nu}+\sum_{q_{i}=d,s,b}\frac{\kappa_{z}^{q_{i}}}{2\Lambda}g_{z}\overline{b}'\sigma_{\mu\nu}q_{i}Z^{\mu\nu}+\sum_{q_{i}=d,s,b}\frac{\kappa_{g}^{q_{i}}}{2\Lambda}g_{s}\overline{b}'\sigma_{\mu\nu}\lambda_{a}q_{i}G_{a}^{\mu\nu}+h.c.\label{eq:eq1}
\end{equation}
where $F^{\mu\nu}$, $Z^{\mu\nu}$ and $G^{\mu\nu}$ are the field
strength tensors of the gauge bosons; $\sigma_{\mu\nu}=i(\gamma_{\mu}\gamma_{\nu}-\gamma_{\nu}\gamma_{\mu})/2$;
$\lambda_{a}$ are the Gell-Mann matrices; $Q_{q}$ is the electric
charge of the quark ($q$); $g_{e}$, $g_{Z}$ and $g_{s}$ are the
electromagnetic, neutral weak and the strong coupling constants, respectively.
$g_{Z}=g_{e}/\cos\theta_{w}\sin\theta_{w}$, where $\theta_{w}$ is
the weak mixing angle. $\kappa_{\gamma}$ is the anomalous coupling
with photon; $\kappa_{z}$ is for the $Z$ boson, and $\kappa_{g}$
is the coupling with gluon. Finally, $\Lambda$ is the cutoff scale
for the new interactions.

\section{Decay Widths and Branch\i ngs}

For the decay channels $Q'\to Vq$ where $V\equiv\gamma,Z,g$, we
use the effective Lagrangian to calculate the anomalous decay widths 

\begin{equation}
\Gamma(Q'\to gq)=\frac{2}{3}\left(\frac{\kappa_{g}^{q}}{\Lambda}\right)^{2}\alpha_{s}m_{Q'}^{3}\lambda_{0}\label{eq:eq2}
\end{equation}

\begin{equation}
\Gamma(Q'\to\gamma q)=\frac{1}{2}\left(\frac{\kappa_{\gamma}^{q}}{\Lambda}\right)^{2}\alpha_{e}Q_{q}^{2}m_{Q'}^{3}\lambda_{0}\label{eq:eq3}
\end{equation}

\begin{equation}
\Gamma(Q'\to Zq)=\frac{1}{16}\left(\frac{\kappa_{Z}^{q}}{\Lambda}\right)^{2}\frac{\alpha_{e}m_{Q'}^{3}}{\sin^{2}\theta_{W}\cos^{2}\theta_{W}}\lambda_{Z}\sqrt{\lambda_{r}}\label{eq:eq:4}
\end{equation}
with

\begin{equation}
\lambda_{0}=1-3m_{q}^{2}/m_{Q'}^{2}+3m_{q}^{4}/m_{Q'}^{4}-m_{q}^{6}/m_{Q'}^{6}\label{eq:eq:5}
\end{equation}

\begin{equation}
\lambda_{r}=1+m_{W}^{4}/m_{Q'}^{4}+m_{q}^{4}/m_{Q'}^{4}-2m_{W}^{2}/m_{Q'}^{2}-2m_{q}^{2}/m_{Q'}^{2}-2m_{W}^{2}m_{q}^{2}/m_{Q'}^{4}\label{eq:6}
\end{equation}

\begin{equation}
\lambda_{Z}=2-m_{Z}^{2}/m_{Q'}^{2}-4m_{q}^{2}/m_{Q'}^{2}+2m_{q}^{4}/m_{Q'}^{4}-6m_{q}m_{Z}^{2}/m_{Q'}^{3}-m_{Z}^{2}m_{t}^{2}/m_{Q'}^{4}-m_{Z}^{4}/m_{Q'}^{4}\label{eq:eq:7}
\end{equation}

The anomalous decay widths in different channels are proportional
to $\Lambda^{-2}$, and they are assumed to be dominant for $\kappa/\Lambda>0.1$
TeV$^{-1}$ over the charged current channels. In this case, if we
take all the anomalous coupling equal then the branching ratios will
be nearly independent of $\kappa/\Lambda$. We have used three parametrizations
sets entitled PI, PII and PIII. For the PI parametrization, we assume
the constant value $\kappa_{i}/\Lambda=0.1$ TeV $^{-1}$, and PII
has the parameters $\kappa_{i}/\Lambda=0.1\lambda^{4-i}$ TeV $^{-1}$
with $\lambda=0.5$. For PIII we take the couplings $\kappa_{i}/\Lambda=0.5\lambda^{4-i}$
TeV$^{-1}$ with the same value of $\lambda$. The index $i$ is the
generation number. 

Table \ref{tab:tab1} and Table \ref{tab:tab2} present the decay
width and branching ratios of the new heavy quark $t'$ through anomalous
interactions for the parametrizations PI, PII and PIII, respectively.
Taking the anomalous coupling $\kappa/\Lambda=0.1$ TeV$^{-1}$ we
calculate the $t'$ decay width $\Gamma=0.65$ GeV and $1.90$ GeV
for $m_{t'}=700$ GeV and $1000$ GeV, respectively. The branching
into $t'\rightarrow qg$ channel is the largest and branching into
$t'\rightarrow q\gamma$ channel is the smallest for equal anomalous
couplings with the parametrization PI. On the other hand, PII and
PIII parametrizations give higher branching ratios into $tV$ ($V=g,\, Z,\,\gamma$)
than $qV$ ($q=u,\, c$$ $) channels due to $\lambda^{4-i}$ factor
in the parametrizations. 

For the new heavy quark $b'$ the decay witdh and branching ratios
are presented in Table \ref{tab:tab3} and Table \ref{tab:tab4} for
the parametrizations PI, PII and PIII, respectively. We calculate
the $b'$ decay width, by taking the anomalous coupling $\kappa/\Lambda=0.1$
TeV$^{-1}$, $\Gamma=0.68$ GeV and $1.92$ GeV for $m_{b'}=700$
GeV and $1000$ GeV, respectively. The branching for $b'\rightarrow qg$
is the largest (30\%) and its the smallest for $b'\rightarrow q\gamma$
(0.2\%) channel for equal anomalous couplings with the parametrization
PI. For PII and PIII parametrizations the branching ratios into $bV$
($V=g,\, Z,\,\gamma$) are larger than $qV$ ($q=d,\, s$) channels.
The $t'$ and $b'$ decay widths are about the same values for PII
and PIII parametrizations.

\begin{table}
\protect\caption{Branching ratios (\%) and decay width of the heavy quarks ($t'$)
with only anomalous interactions for PI parametrization and $\kappa/\Lambda=0.1$
TeV$^{-1}$. \label{tab:tab1}}

\begin{tabular}{|c|c|c|c|c|c|c|c|}
\hline 
Mass(GeV) & $gu(c)$ & $gt$ & $Zu(c)$ & $Zt$ & $\gamma u(c)$ & $\gamma t$ & $\Gamma$(GeV)\tabularnewline
\hline 
\hline 
$500$ & 33.5 & 22.9 & 2.86 & 1.82 & 0.92 & 0.63 & 0.23\tabularnewline
\hline 
$600$ & 32.3 & 25.0 & 2.86 & 2.13 & 0.91 & 0.70 & 0.41\tabularnewline
\hline 
$700$ & 31.6 & 26.2 & 2.87 & 2.34 & 0.90 & 0.75 & 0.65\tabularnewline
\hline 
$800$ & 31.1 & 27.0 & 2.89 & 2.48 & 0.90 & 0.78 & 0.97\tabularnewline
\hline 
$900$ & 30.7 & 27.5 & 2.91 & 2.58 & 0.91 & 0.81 & 1.39\tabularnewline
\hline 
$1000$ & 30.5 & 27.8 & 2.93 & 2.66 & 0.91 & 0.83 & 1.90\tabularnewline
\hline 
\end{tabular}
\end{table}

\begin{table}
\protect\caption{The same as Table \ref{tab:tab1}, but for PII (PIII) parametrizations.
\label{tab:tab2}}

\begin{tabular}{|c|c|c|c|c|c|c|c|c|c|c|}
\hline 
Mass(GeV) & $gu$ & $gc$ & $gt$ & $Zu$ & $Zc$ & $Zt$ & $\gamma u$ & $\gamma c$ & $\gamma t$ & $\Gamma$(GeV)\tabularnewline
\hline 
\hline 
$500$ & 5.66 & 22.60 & 61.90 & 0.48 & 1.93 & 4.92 & 0.15 & 0.62 & 1.71 & 0.021 (0.558)\tabularnewline
\hline 
$600$ & 5.17 & 20.70 & 63.90 & 0.46 & 1.83 & 5.46 & 0.14 & 0.58 & 1.80 & 0.040 (1.024)\tabularnewline
\hline 
$700$ & 4.90 & 19.60 & 64.90 & 0.44 & 1.78 & 5.79 & 0.14 & 0.56 & 1.87 & 0.066 (1.68)\tabularnewline
\hline 
$800$ & 4.73 & 18.90 & 65.60 & 0.44 & 1.76 & 6.02 & 0.14 & 0.55 & 1.91 & 0.100 (2.561)\tabularnewline
\hline 
$900$ & 4.61 & 18.40 & 65.90 & 0.44 & 1.74 & 6.19 & 0.13 & 0.54 & 1.95 & 0.145 (3.680)\tabularnewline
\hline 
$1000$ & 4.53 & 18.10 & 66.20 & 0.43 & 1.74 & 6.32 & 0.13 & 0.54 & 1.98 & 0.200 (5.070)\tabularnewline
\hline 
\end{tabular}
\end{table}

\begin{table}
\protect\caption{Branching ratios (\%) and decay width of the heavy quarks ($b'$)
with only anomalous interactions for PI parametrization and $\kappa/\Lambda=0.1$
TeV$^{-1}$. \label{tab:tab3}}

\begin{tabular}{|c|c|c|c|c|}
\hline 
Mass(GeV) & $gd(s,b)$ & $Zd(s,b)$ & $\gamma d(s,b)$ & $\Gamma$(GeV)\tabularnewline
\hline 
\hline 
$500$ & 30.50 & 2.60 & 0.21 & 0.257\tabularnewline
\hline 
$600$ & 30.40 & 2.69 & 0.21 & 0.436\tabularnewline
\hline 
$700$ & 30.40 & 2.76 & 0.22 & 0.682\tabularnewline
\hline 
$800$ & 30.30 & 2.82 & 0.22 & 1.005\tabularnewline
\hline 
$900$ & 30.20 & 2.86 & 0.22 & 1.415\tabularnewline
\hline 
$1000$ & 30.20 & 2.90 & 0.23 & 1.921\tabularnewline
\hline 
\end{tabular}
\end{table}

\begin{table}
\protect\caption{The same as Table \ref{tab:tab3}, but for PII (PIII) parametrizations.
\label{tab:tab4}}

\begin{tabular}{|c|c|c|c|c|c|c|c|c|c|c|}
\hline 
Mass(GeV) & $gd$ & $gs$ & $gb$ & $Zd$ & $Zs$ & $Zb$ & $\gamma d$ & $\gamma s$ & $\gamma b$ & $\Gamma$(GeV)\tabularnewline
\hline 
\hline 
$500$ & 4.36 & 17.40 & 69.80 & 0.37 & 1.49 & 5.95 & 0.030 & 0.12 & 0.48 & 0.028 (0.704)\tabularnewline
\hline 
$600$ & 4.35 & 17.40 & 69.50 & 0.38 & 1.54 & 6.16 & 0.030 & 0.12 & 0.49 & 0.047 (1.194)\tabularnewline
\hline 
$700$ & 4.34 & 17.30 & 69.40 & 0.39 & 1.58 & 6.31 & 0.031 & 0.12 & 0.50 & 0.074 (1.866)\tabularnewline
\hline 
$800$ & 4.33 & 17.30 & 69.20 & 0.40 & 1.61 & 6.44 & 0.031 & 0.12 & 0.50 & 0.110 (2.749)\tabularnewline
\hline 
$900$ & 4.32 & 17.30 & 69.10 & 0.41 & 1.64 & 6.54 & 0.032 & 0.13 & 0.51 & 0.154 (3.869)\tabularnewline
\hline 
$1000$ & 4.32 & 17.30 & 69.00 & 0.41 & 1.66 & 6.63 & 0.032 & 0.13 & 0.52 & 0.210 (5.253)\tabularnewline
\hline 
\end{tabular}
\end{table}

\section{The Cross Sect\i ons}

\begin{figure}
\includegraphics[scale=0.8]{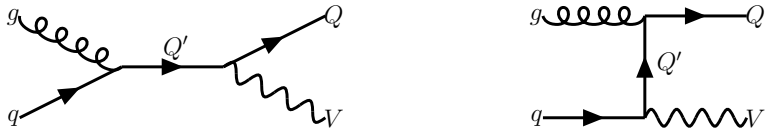}

\protect\caption{Diagrams for the subprocess $gq\to VQ$ with anomalous vertices $Q'qV$
and $Q'QV$ (where $Q'$ can be the heavy quark $b'$ or $t'$ depending
on the type of light ($q$) or heavy $(Q\equiv t,\, b)$ quarks, respectively).
\label{fig:fig1}}
\end{figure}

In order to study the new heavy quark productions at the LHC, we have
used effective anomalous interaction vertices and implemented these
vertices into the CalcHEP package \cite{CalcHEP}. In all of the numerical
calculations, the parton distribution function are set to the CTEQ6L
parametrization \cite{CTEQ6L}. The new heavy quarks can be produced
through its anomalous couplings to the ordinary quarks and neutral
vector bosons as shown in Fig. \ref{fig:fig1}. 

\begin{figure}
\includegraphics[scale=0.8]{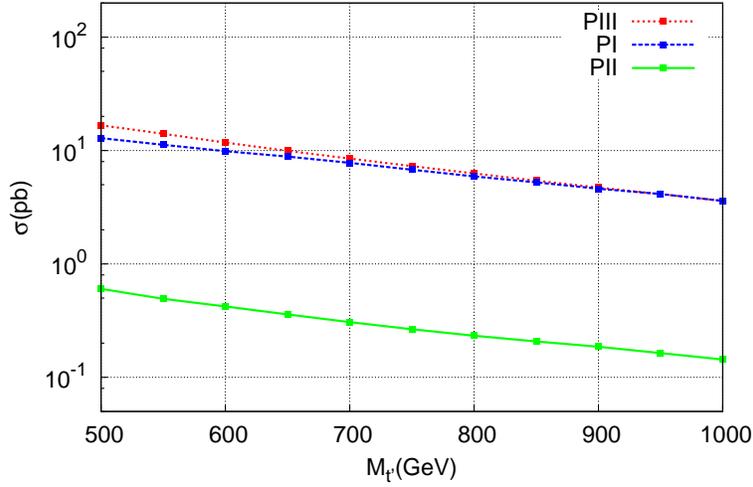}\protect\caption{The cross section for the process $pp\rightarrow tV+X$ depending
on the mass for parameter sets PI, PII and PIII at the center of mass
energy $\sqrt{s}=13$ TeV. \label{fig:fig:2}}
\end{figure}

\begin{figure}
\includegraphics[scale=0.8]{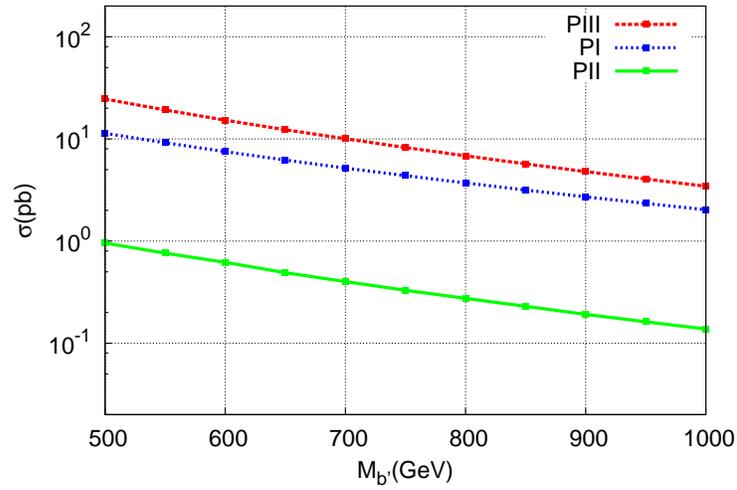}\protect\caption{The cross section for the process $pp\rightarrow bX$ depending on
the new heavy quark mass for parameter sets PI, PII and PII at the
center of mass energy $\sqrt{s}=13$ TeV. \label{fig:fig:3}}
\end{figure}

Total cross sections for the productions of new heavy quarks $t'$
and $b'$ are given in Table \ref{tab:tab5} and Table \ref{tab:tab6}
for the parametrizations PI, PII and PIII, at the center of mass energy
of 8 TeV and 13 TeV. For an illustration, taking the mass of new heavy
quarks as 700 GeV the cross section of $t'(b')$ production is calculated
as 8.50 pb (10.03 pb) for the parametrization PIII at $\sqrt{s}=13$
TeV. It can be seen from Table \ref{tab:tab5} and Table \ref{tab:tab6},
the cross sections decreases while the mass of the new heavy quark
increases. The cross section for $t'$ production is larger than the
$b'$ production with a factor of 1.2-1.8 (0.7-1.0) for PI (PII and
PIII) parametrization depending on the considered mass range at $\sqrt{s}=13$
TeV. The general behaviours of the production cross sections depending
on the mass of heavy quarks are presented in Fig. \ref{fig:fig:2}
and Fig. \ref{fig:fig:3} for different parametrizations.

\begin{table}
\protect\caption{The cross sections (in pb) of heavy quark $t'$ production without
cuts for PI, PII and PIII parametrizations at the center of mass energy
$13$ TeV ($8$ TeV), respectively. \label{tab:tab5}}

\begin{tabular}{|c|c|c|c|}
\hline 
\multirow{2}{*}{Mass (GeV)} & PI  & PII & PIII\tabularnewline
\cline{2-4} 
 & $\sqrt{s}=$13 TeV (8 TeV) & $\sqrt{s}=$13 TeV (8 TeV) & $\sqrt{s}=$13 TeV (8 TeV)\tabularnewline
\hline 
\hline 
$500$ & 13.733 (5.30) & 0.664 (0.244) & 16.736 (6.113)\tabularnewline
\hline 
$600$ & 10.362(3.72) & 0.464 (0.159) & 11.770 (4.031)\tabularnewline
\hline 
$700$ & 7.825 (2.64) & 0.337 (0.109) & 8.502 (2.718)\tabularnewline
\hline 
$800$ & 5.961 (1.89) & 0.250 (0.075) & 6.276 (1.882)\tabularnewline
\hline 
$900$ & 4.602 (1.36) & 0.189 (0.053) & 4.701 (1.326)\tabularnewline
\hline 
$1000$ & 3.593 (0.98) & 0.144 (0.038) & 3.609 (0.950)\tabularnewline
\hline 
\end{tabular}
\end{table}

\begin{table}
\protect\caption{The cross sections (in pb) of heavy quark $b'$ production without
cuts for PI, PII and PIII parametrizations at the center of mass energy
of $13$ TeV ($8$ TeV), respectively. \label{tab:tab6}}

\begin{tabular}{|c|c|c|c|}
\hline 
\multirow{2}{*}{Mass (GeV)} & \multicolumn{1}{c|}{PI} & PII & PIII\tabularnewline
\cline{2-4} 
 & $\sqrt{s}=$13 TeV (8 TeV) & $\sqrt{s}=$13 TeV (8 TeV) & $\sqrt{s}=$13 TeV (8 TeV)\tabularnewline
\hline 
\hline 
$500$ & 11.340 (3.913) & 0.970 (0.285) & 24.474 (7.114)\tabularnewline
\hline 
$600$ & 7.495 (2.410) & 0.607 (0.162) & 15.290 (4.09)\tabularnewline
\hline 
$700$ & 5.179 (1.546) & 0.412 (0.099) & 10.031 (2.483)\tabularnewline
\hline 
$800$ & 3.697 (1.025) & 0.286 (0.062) & 6.832 (1.566)\tabularnewline
\hline 
$900$ & 2.707 (0.697) & 0.1905 (0.040) & 4.791 (1.018)\tabularnewline
\hline 
$1000$ & 2.021 (0.482) & 0.137 (0.027) & 3.441 (0.678)\tabularnewline
\hline 
\end{tabular}
\end{table}

\subsection{Analysis of the process $pp\to W^{+}bV+X$ ($V=g,\, Z,\,\gamma$)
for $t'$ signal}

The signal process $pp\to W^{+}bV+X$ ($V=g,\, Z,\,\gamma$) includes
the $t'$ exchange both in the $s$-channel and $t$-channel. The
$s$-channel contribution to the signal process would appear itself
as resonance around the $t'$ mass value in the $WbV$ invariant mass.
The $t$-channel gives the non-resonant contribution. We consider
that the $W$ boson decays into lepton+missing transverse momentum
with the branching ratio $21\%$ and $Z$ boson decays into dilepton
with the branching $6.7\%$. In our analyses, we consider the $t'$
signal in the $l+b_{jet}+\gamma+MET$ , $ $$l+b_{jet}+j+MET$ and
$3l+b_{jet}+MET$ channels, where $l=e,\,\mu$. However, if one takes
the hadronic $W$decays the signal will be enhanced by a factor of
$BR(W\rightarrow\mbox{hadrons})/BR(W\rightarrow l\nu).$ 

We have obtained the cross sections by using the cuts pseudorapidity
$|\eta_{j,\gamma}|<2.5$ and transverse momentum $p_{T}^{j,\gamma}>20-200$
GeV for jets and photon, in Table \ref{tab:tab7} (Table\ref{tab:tab8},
Table \ref{tab:tab9}) for PI (PII, PIII) parametrizations, respectively.
It appears from signal significance calculations that the optimized
transverse momentum cut is $p_{T}$ >100 GeV for $t'$ analyses. 

The backgrounds for the final state $W^{+}b(\bar{b})V$ (where $V\equiv$
photon, jet and $Z$ boson) are given in Table \ref{tab:tab10}. We
apply the following cuts to the final state photon and jets as $|\eta_{j,\gamma}|<2.5$
and $p_{T}^{j,\gamma}>20-200$ GeV. For the background cross section
estimates, we assume the efficiency for \emph{b-}tagging to be $\varepsilon_{b}=50\%$,
and the rejection ratios $10\%$ for $c\,(\bar{c})$ quark jets and
$1\%$ for light quark jets since they are assumed to be mistagged
as $b$-jets. 

\begin{figure}
\includegraphics{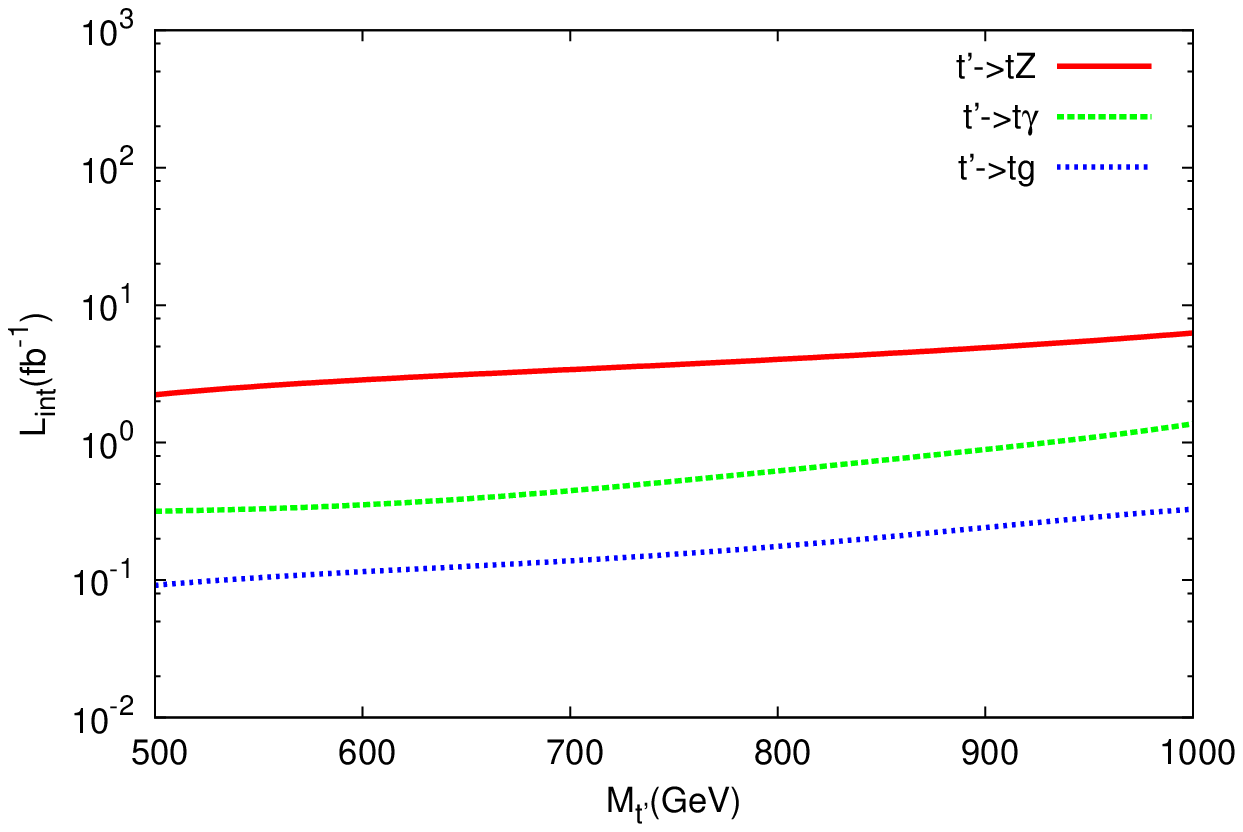}\protect\caption{Integrated luminosity required to reach $3\sigma$ significance for
the signal of $t'$ anomalous interactions for parametrization PI
at the LHC with $\sqrt{s}=13$ TeV. \label{fig:fig4}}
\end{figure}

\begin{figure}
\includegraphics{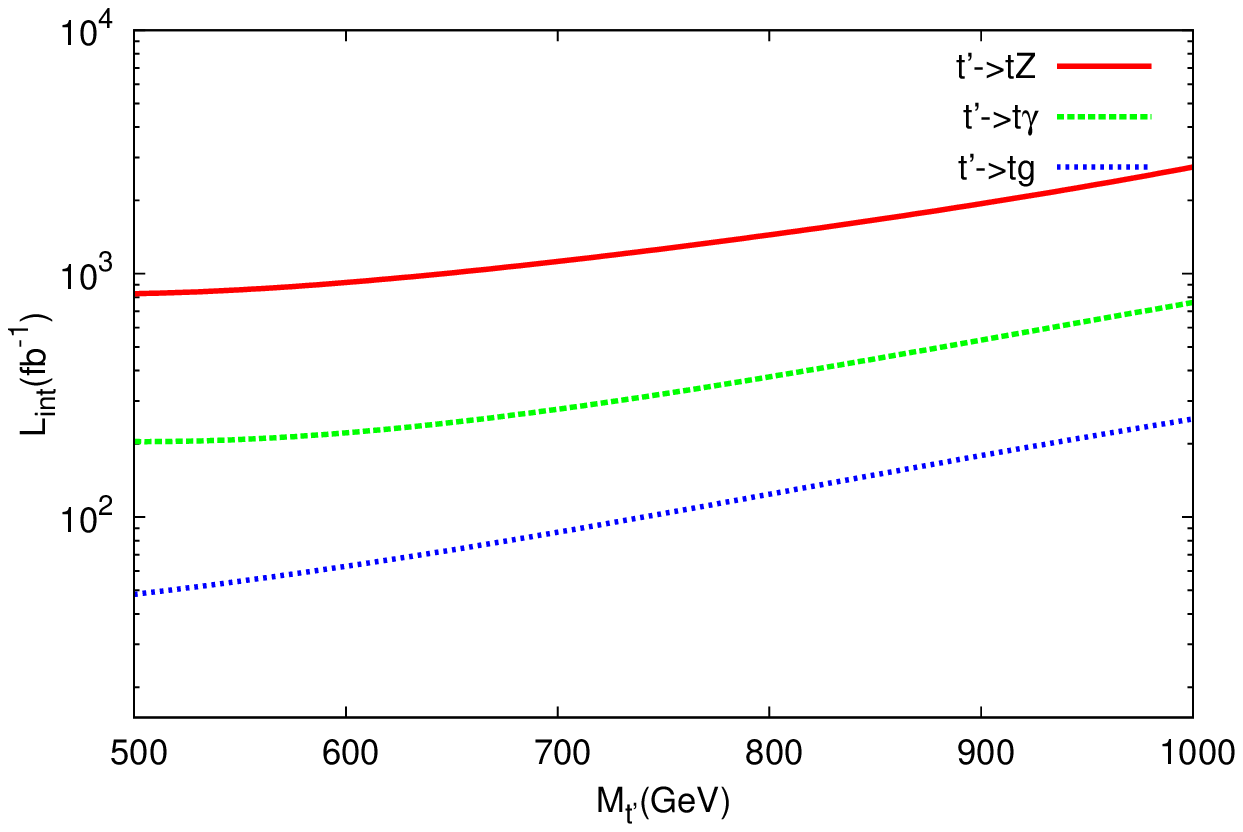}

\protect\caption{The same as Fig.\ref{fig:fig4}, but for parametrization PII.\label{fig:fig5}}
\end{figure}

\begin{figure}
\includegraphics{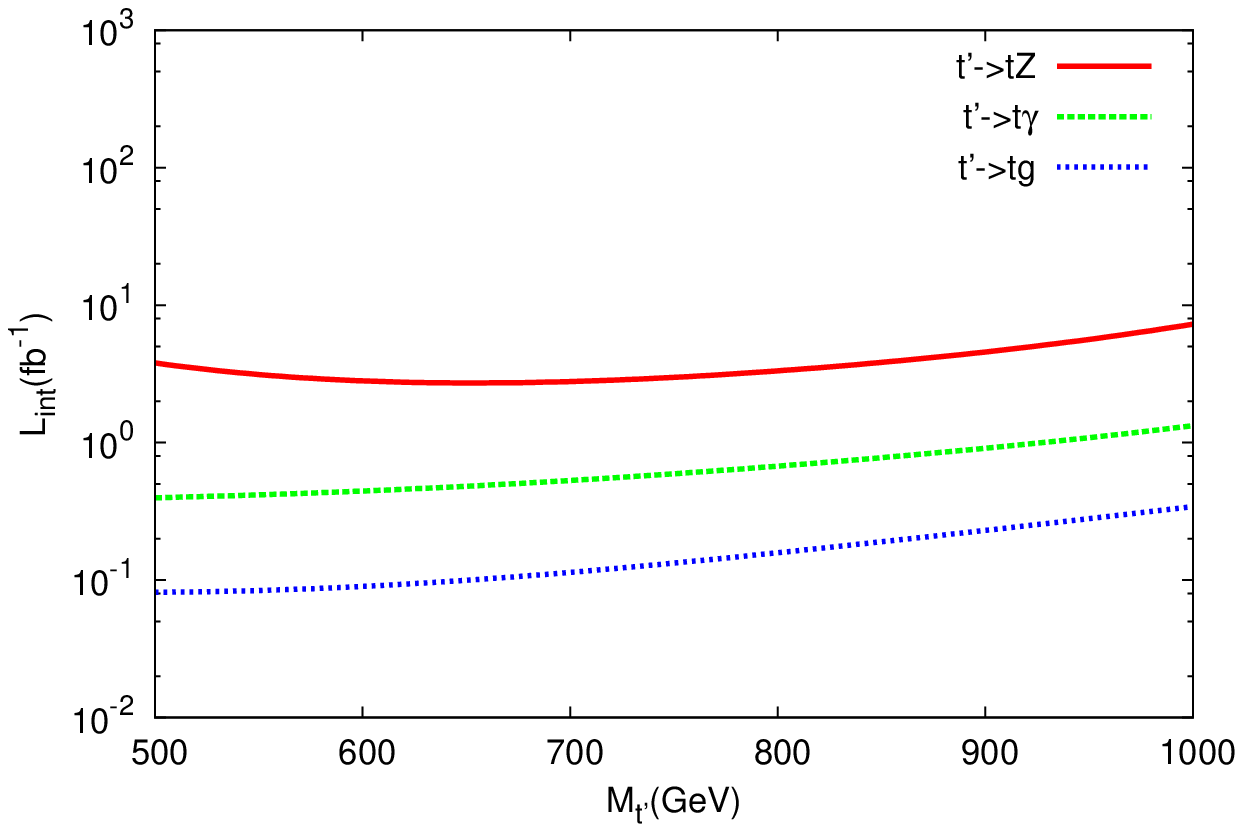}\protect\caption{The same as Fig. \ref{fig:fig4}, but for parametrization PIII.\label{fig:fig6}}
\end{figure}

\begin{figure}
\includegraphics{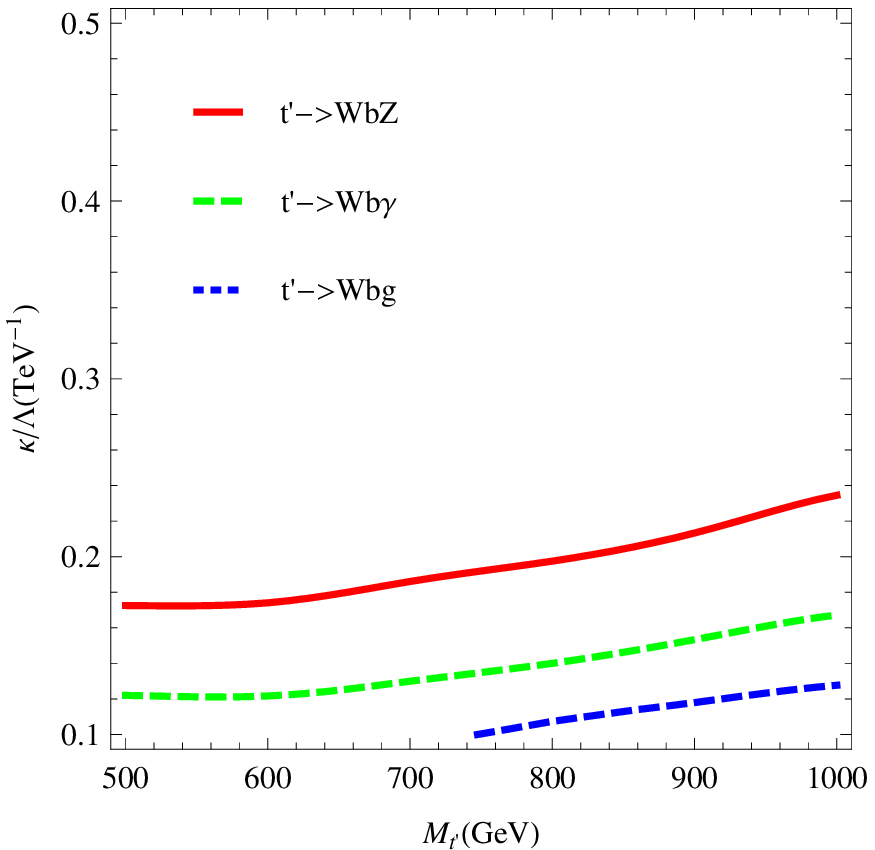}

\protect\caption{The contour plot of anomalous coupling and mass of new heavy quark
$t'$ for the dynamical parametrization explained in the text with
a significance $3\sigma$ at $\sqrt{s}=13$ TeV and $L_{int}=100$
fb$^{-1}$. \label{fig:fig7}}
\end{figure}

In order to find the discovery limits we use the statistical significance
as 

\begin{equation}
SS=\sqrt{2\left[(S+B)\ln(1+\frac{S}{B})-S\right]}\label{eq:8}
\end{equation}
where $S$ and $B$ are the numbers of the signal and background events,
respectively. In Figs. \ref{fig:fig4}- \ref{fig:fig6}, the integrated
luminosity required to reach $3\sigma$ significance for the signal
of $t'$ anomalous interactions is shown for parametrization PI, PII
and PIII at the LHC with $\sqrt{s}=13$ TeV. It is seen from these
figures that the channel $t'\rightarrow tZ$ requires more integrated
luminosity than the other channels. By requiring the signal significance
$SS=3$, the contour plots of $\kappa/\Lambda$ and mass of $t'$quark
are presented in Fig. \ref{fig:fig7}. The results show that one can
discover the $t'$ quark anomalous couplings $\kappa/\Lambda$ down
to 0.1 TeV $^{-1}$ in the $tg$ channel for $m_{t'}$=750 GeV. 

\begin{table}
\protect\caption{The cross sections (in pb) for $t'$ signal in different decay channels
for PI parametrization with $p_{T}$ cuts on the jets and photon and
$|\eta_{j,\gamma}|<2.5$ at the center of mass energy $\sqrt{s}=13$
TeV. \label{tab:tab7}}

\begin{tabular}{|c|c|c|c|c|}
\hline 
Signal & \multicolumn{4}{c|}{PI}\tabularnewline
\hline 
$pp\rightarrow W^{+}b\gamma X$ & $p_{T}>20$ GeV & $p_{T}>50$ GeV & $p_{T}>100$ GeV & $p_{T}>200$ GeV\tabularnewline
\hline 
\hline 
$500$ & $2.89\times10^{-1}$  & $2.10\times10^{-1}$  & $1.24\times10^{-1}$  & $1.02\times10^{-4}$\tabularnewline
\hline 
$600$ & $2.43\times10^{-1}$  & $1.64\times10^{-1}$  & $1.19\times10^{-1}$ & $1.23\times10^{-2}$\tabularnewline
\hline 
$700$ & $1.68\times10^{-1}$ & $1.2\times10^{-1}$  & $1.12\times10^{-1}$  & $2.25\times10^{-2}$ \tabularnewline
\hline 
$800$ & $1.30\times10^{-1}$  & $1.03\times10^{-1}$  & $7.53\times10^{-2}$  & $3.25\times10^{-2}$ \tabularnewline
\hline 
$900$ & $1.02\times10^{-1}$  & $8.08\times10^{-2}$  & $6.96\times10^{-2}$  & $3.02\times10^{-2}$ \tabularnewline
\hline 
$1000$ & $7.61\times10^{-2}$  & $6.35\times10^{-2}$  & $5.07\times10^{-2}$  & $2.94\times10^{-2}$ \tabularnewline
\hline 
$pp\rightarrow W^{+}bgX$ &  &  &  & \tabularnewline
\hline 
$500$ & $7.78\times10^{0}$  & $6.02\times10^{0}$  & $3.63\times10^{0}$  & $4.74\times10^{-3}$\tabularnewline
\hline 
$600$ & $6.30\times10^{0}$ & $5.18\times10^{0}$  & $3.13\times10^{0}$  & $2.58\times10^{-1}$\tabularnewline
\hline 
$700$ & $4.99\times10^{0}$  & $3.63\times10^{0}$  & $3.04\times10^{0}$  & $9.32\times10^{-1}$\tabularnewline
\hline 
$800$ & $4.01\times10^{0}$  & $3.45\times10^{0}$  & $2.76\times10^{0}$  & $9.91\times10^{-1}$\tabularnewline
\hline 
$900$ & $3.32\times10^{0}$  & $2.77\times10^{0}$  & $2.13\times10^{0}$  & $1.08\times10^{0}$\tabularnewline
\hline 
$1000$ & $2.58\times10^{0}$  & $2.27\times10^{0}$  & $1.88\times10^{0}$ & $1.01\times10^{0}$\tabularnewline
\hline 
$pp\rightarrow W^{+}bZX$ &  &  &  & \tabularnewline
\hline 
$500$ & $7.96\times10^{-1}$ & $6.01\times10^{-1}$ & $3.01\times10^{-1}$  & $1.01\times10^{-4}$ \tabularnewline
\hline 
$600$ & $4.79\times10^{-1}$ & $3.86\times10^{-1}$  & $2.45\times10^{-1}$  & $2.71\times10^{-3}$\tabularnewline
\hline 
$700$ & $3.99\times10^{-1}$  & $3.12\times10^{-1}$  & $2.39\times10^{-1}$ & $6.96\times10^{-2}$ \tabularnewline
\hline 
$800$ & $3.31\times10^{-1}$  & $2.89\times10^{-1}$  & $2.09\times10^{-1}$  & $8.05\times10^{-2}$\tabularnewline
\hline 
$900$ & $2.73\times10^{-1}$  & $2.73\times10^{-1}$  & $1.91\times10^{-1}$  & $9.54\times10^{-2}$\tabularnewline
\hline 
$1000$ & $2.23\times10^{-1}$  & $2.02\times10^{-1}$  & $1.61\times10^{-1}$  & $9.10\times10^{-2}$\tabularnewline
\hline 
\end{tabular}
\end{table}

\begin{table}
\protect\caption{The same as Table VII, but for parametrization PII. \label{tab:tab8}}

\begin{tabular}{|c|c|c|c|c|}
\hline 
Signal & \multicolumn{4}{c|}{PII}\tabularnewline
\hline 
$pp\rightarrow W^{+}b\gamma X$ & $p_{T}>20$ GeV & $p_{T}>50$ GeV & $p_{T}>100$ GeV & $p_{T}>200$ GeV\tabularnewline
\hline 
\hline 
$500$ & $6.78\times10^{-3}$  & $5.07\times10^{-3}$  & $3.45\times10^{-3}$  & $2.64\times10^{-7}$\tabularnewline
\hline 
$600$ & $6.57\times10^{-3}$ & $5.42\times10^{-3}$  & $3.47\times10^{-3}$  & $5.34\times10^{-4}$\tabularnewline
\hline 
$700$ & $5.02\times10^{-3}$  & $4.31\times10^{-3}$  & $3.04\times10^{-3}$  & $8.73\times10^{-4}$\tabularnewline
\hline 
$800$ & $3.91\times10^{-3}$  & $3.76\times10^{-3}$  & $2.56\times10^{-3}$ &  $1.03\times10^{-3}$\tabularnewline
\hline 
$900$ & $3.03\times10^{-3}$  & $2.68\times10^{-3}$  & $2.11\times10^{-3}$  & $1.01\times10^{-3}$\tabularnewline
\hline 
$1000$ & $2.40\times10^{-3}$  & $2.43\times10^{-3}$  & $1.77\times10^{-3}$  & $9.98\times10^{-4}$\tabularnewline
\hline 
$pp\rightarrow W^{+}bgX$ &  &  &  & \tabularnewline
\hline 
$500$ & $3.47\times10^{-1}$  & $2.68\times10^{-1}$  & $1.52\times10^{-1}$ & $5.30\times10^{-6}$\tabularnewline
\hline 
$600$ & $2.51\times10^{-1}$  & $2.12\times10^{-1}$  & $1.35\times10^{-1}$ & $2.01\times10^{-2}$\tabularnewline
\hline 
$700$ & $1.87\times10^{-1}$  & $1.6\times10^{-1}$  & $1.16\times10^{-1}$  & $3.42\times10^{-2}$\tabularnewline
\hline 
$800$ & $1.46\times10^{-1}$ & $1.25\times10^{-1}$  & $9.39\times10^{-2}$  & $4.03\times10^{-2}$\tabularnewline
\hline 
$900$ & $1.12\times10^{-1}$  & $1.08\times10^{-1}$  & $7.80\times10^{-2}$  & $3.86\times10^{-2}$\tabularnewline
\hline 
$1000$ & $9.35\times10^{-2}$  & $8.37\times10^{-2}$  & $6.62\times10^{-2}$  & $3.68\times10^{-2}$\tabularnewline
\hline 
$pp\rightarrow W^{+}bZX$ &  &  &  & \tabularnewline
\hline 
$500$ & $2.10\times10^{-2}$ & $1.77\times10^{-2}$ & $1.16\times10^{-2}$ &  $2.64\times10^{-7}$\tabularnewline
\hline 
$600$ & $1.95\times10^{-2}$  & $1.75\times10^{-2}$  & $1.14\times10^{-2}$  & $1.34\times10^{-3}$\tabularnewline
\hline 
$700$ & $1.73\times10^{-2}$  & $1.43\times10^{-2}$  & $1.00\times10^{-2}$  & $2.9\times10^{-3}$\tabularnewline
\hline 
$800$ & $1.34\times10^{-2}$  & $1.19\times10^{-2}$  & $8.89\times10^{-3}$  & $3.42\times10^{-3}$\tabularnewline
\hline 
$900$ & $1.06\times10^{-2}$ & $9.55\times10^{-3}$  & $7.63\times10^{-3}$  & $3.37\times10^{-3}$\tabularnewline
\hline 
$1000$ & $8.09\times10^{-3}$  & $7.58\times10^{-3}$  & $6.31\times10^{-3}$  & $3.23\times10^{-3}$\tabularnewline
\hline 
\end{tabular}
\end{table}

\begin{table}
\protect\caption{The same as Table \ref{tab:tab7}, but for parametrization PIII. \label{tab:tab9} }

\begin{tabular}{|c|c|c|c|c|}
\hline 
Signal & \multicolumn{4}{c|}{PIII}\tabularnewline
\hline 
$pp\rightarrow W^{+}b\gamma X$ & $p_{T}>20$ GeV & $p_{T}>50$ GeV & $p_{T}>100$ GeV & $p_{T}>200$ GeV\tabularnewline
\hline 
\hline 
$500$ & $2.60\times10^{-1}$ & $2.78\times10^{-1}$ & $1.08\times10^{-1}$ & $1.59\times10^{-4}$\tabularnewline
\hline 
$600$ & $1.78\times10^{-1}$ & $1.61\times10^{-1}$ & $1.01\times10^{-1}$ & $1.42\times10^{-2}$\tabularnewline
\hline 
$700$ & $1.56\times10^{-1}$ & $1.35\times10^{-1}$ & $9.33\times10^{-2}$ & $2.72\times10^{-2}$\tabularnewline
\hline 
$800$ & $1.17\times10^{-1}$ & $1.06\times10^{-1}$ & $7.84\times10^{-2}$ & $3.32\times10^{-2}$\tabularnewline
\hline 
$900$ & $9.04\times10^{-2}$ & $8.42\times10^{-2}$ & $6.68\times10^{-2}$ & $3.25\times10^{-2}$\tabularnewline
\hline 
$1000$ & $ $$7.6\times10^{-2}$ & $6.76\times10^{-2}$ & $5.16\times10^{-2}$ & $3.17\times10^{-2}$\tabularnewline
\hline 
$pp\rightarrow W^{+}bgX$ &  &  &  & \tabularnewline
\hline 
$500$ & $8.39\times10^{0}$ & $6.49\times10^{0}$ & $3.86\times10^{0}$ & $4.65\times10^{-3}$\tabularnewline
\hline 
$600$ & $6.10\times10^{0}$ & $5.78\times10^{0}$ & $3.81\times10^{0}$ & $5.56\times10^{-1}$\tabularnewline
\hline 
$700$ & $5.39\times10^{0}$ & $4.64\times10^{0}$ & $3.41\times10^{0}$ & $9.70\times10^{-1}$\tabularnewline
\hline 
$800$ & $3.94\times10^{0}$ & $3.54\times10^{0}$ & $2.73\times10^{0}$ & $1.05\times10^{0}$\tabularnewline
\hline 
$900$ & $3.24\times10^{0}$ & $2.76\times10^{0}$ & $2.27\times10^{0}$ & $1.07\times10^{0}$\tabularnewline
\hline 
$1000$ & $2.33\times10^{0}$ & $2.29\times10^{0}$ & $1.84\times10^{0}$ & $9.98\times10^{-1}$\tabularnewline
\hline 
$pp\rightarrow W^{+}bZX$ &  &  &  & \tabularnewline
\hline 
$500$ & $7.72\times10^{-1}$ & $1.01\times10^{0}$ & $2.17\times10^{-1}$ & $6.27\times10^{-4}$\tabularnewline
\hline 
$600$ & $6.24\times10^{-1}$ & $3.85\times10^{-1}$ & $2.92\times10^{-1}$ & $3.20\times10^{-2}$\tabularnewline
\hline 
$700$ & $5.00\times10^{-1}$$ $ & $3.05\times10^{-1}$ & $2.86\times10^{-1}$ & $5.80\times10^{-2}$\tabularnewline
\hline 
$800$ & $3.78\times10^{-1}$ & $2.50\times10^{-1}$ & $2.42\times10^{-1}$ & $9.64\times10^{-2}$\tabularnewline
\hline 
$900$ & $3.04\times10^{-1}$ & $1.67\times10^{-1}$ & $2.06\times10^{-1}$ & $9.62\times10^{-2}$\tabularnewline
\hline 
$1000$ & $2.51\times10^{-1}$ & $1.29\times10^{-1}$ & $1.48\times10^{-1}$ & $9.61\times10^{-2}$\tabularnewline
\hline 
\end{tabular}
\end{table}

\begin{table}
\protect\caption{The cross sections (in pb) for the relevant backgrounds ($W^{+}b(\bar{b})V$,
$W^{+}c(\bar{c})V$ and $W^{+}jV$, where $V=$photon, jet and $Z$
boson) with $p_{T}$ cuts on the jets at the center of mass energy
$\sqrt{s}=13$ TeV. \label{tab:tab10}}

\begin{tabular}{|c|c|c|c|c|}
\hline 
Background & $p_{T}>20$ GeV & $p_{T}>50$ GeV & $p_{T}>100$ GeV & $p_{T}>200$ GeV\tabularnewline
\hline 
\hline 
$pp\to W^{+}b\gamma$ & $2.37\times10^{-3}$ & $3.62\times10^{-4}$ & $6.17\times10^{-5}$ & $6.99\times10^{-6}$\tabularnewline
\hline 
$pp\to W^{+}\bar{c}\gamma$ & $4.15\times10^{0}$ & $4.59\times10^{-1}$ & $6.25\times10^{-2}$ & $6.21\times10^{-3}$\tabularnewline
\hline 
$pp\to W^{+}j\gamma$ & $2.63\times10^{1}$ & $4.30\times10^{0}$ & $7.33\times10^{-1}$ & $1.27\times10^{-1}$\tabularnewline
\hline 
$pp\to W^{+}b(\bar{b})j$ & $7.26\times10^{1}$ & $3.02\times10^{1}$ & $6.11\times10^{0}$ & $9.74\times10^{-1}$\tabularnewline
\hline 
$pp\to W^{+}c(\bar{c})j$ & $5.98\times10^{2}$ & $9.65\times10^{1}$ & $1.79\times10^{1}$ & $2.42\times10^{0}$\tabularnewline
\hline 
$pp\to W^{+}jj$ & $7.31\times10^{3}$ & $7.78\times10^{2}$ & $1.61\times10^{2}$ & $2.58\times10^{1}$\tabularnewline
\hline 
$pp\to W^{+}bZ$ & $6.26\times10^{-4}$ & $3.99\times10^{-4}$ & $1.93\times10^{-4}$ & $4.71\times10^{-5}$\tabularnewline
\hline 
$pp\to W^{+}\bar{c}Z$ & $5.29\times10^{-1}$ & $3.40\times10^{-1}$ & $1.66\times10^{-1}$ & $4.15\times10^{-2}$\tabularnewline
\hline 
$pp\to W^{+}jZ$ & $8.59\times10^{0}$ & $4.83\times10^{0}$ & $2.49\times10^{0}$ & $7.91\times10^{-1}$\tabularnewline
\hline 
\end{tabular}
\end{table}

\subsection{Analysis of the process $pp\rightarrow bV+X$ ($V=g,\, Z,\,\gamma$)
for $b'$ signal}

The signal process $pp\to bV+X$ ($V=g,\, Z,\,\gamma$) includes the
new heavy quark $b'$ exchange both in the $s$-channel and $t$-channel.
The $s$- channel contributes to the signal process as resonance around
the $b'$ mass value in the $bV$ invariant mass, while the $t$-
channel contributes to the non-resonant behaviour. For this process,
we consider the leptonic decays of $Z$ boson. In the analyses, we
consider the $b'$ signal to be $b_{jet}+\gamma$ , $ $$b_{jet}+j$
and $b_{jet}+\mbox{dilepton}$. 

We have obtained the cross sections by using the pseudorapidity cuts
$|\eta_{j,\gamma}|<2.5$ and transverse momentum cuts $p_{T}^{j,\gamma}>20-200$
GeV for jets and photon, in Table \ref{tab:tab11} (Table \ref{tab:tab12},
Table \ref{tab:tab13} for PI (PII, PIII) parametrizations, respectively.
It appears from signal significance calculation that the optimized
transverse momentum cut is $p_{T}$ >200 GeV for $b'$ analyses. 

The backgrounds for the final state $b(\bar{b})V$ (where $V=$photon,
jet and $Z$ boson) are given in \ref{tab:tab14}. We apply the following
cuts to the final state photon and jets as $|\eta_{j,\gamma}|<2.5$
and $p_{T}^{j,\gamma}>20-200$ GeV. It can be noted that the background
cross section decreases as the $p_{T}$ cuts increases. We assume
the efficiency for \emph{b-}tagging to be $\varepsilon_{b}=50\%$,
and the rejection ratios $10\%$ for $c\,(\bar{c})$ quark jets and
$1\%$ for light quark jets. 

\begin{table}
\protect\caption{The cross sections (in pb) for $b'$ signal in different decay channel
for parametrization PI with $p_{T}$ cuts on the jets and photon and
$|\eta_{j,\gamma}|<2.5$ at the center of mass energy $\sqrt{s}=13$
TeV. \label{tab:tab11}}

\begin{tabular}{|c|c|c|c|c|}
\hline 
Signal & \multicolumn{4}{c|}{PI}\tabularnewline
\hline 
$pp\rightarrow b\gamma X$ & $p_{T}>20$ GeV & $p_{T}>50$ GeV & $p_{T}>100$ GeV & $p_{T}>200$ GeV\tabularnewline
\hline 
\hline 
$500$ & $5.64\times10^{-2}$ & $5.62\times10^{-2}$ & $5.49\times10^{-2}$ & $3.95\times10^{-2}$\tabularnewline
\hline 
$600$ & $3.96\times10^{-2}$ & $3.96\times10^{-2}$  & $3.90\times10^{-2}$ & $3.33\times10^{-2}$\tabularnewline
\hline 
$700$ & $2.87\times10^{-2}$ & $2.87\times10^{-2}$ & $2.86\times10^{-2}$ & $2.59\times10^{-2}$\tabularnewline
\hline 
$800$ & $2.12\times10^{-2}$ & $2.13\times10^{-2}$ & $2.12\times10^{-2}$ & $1.99\times10^{-2}$\tabularnewline
\hline 
$900$ & $1.60\times10^{-2}$ & $1.60\times10^{-2}$ & $1.60\times10^{-2}$ & $1.53\times10^{-2}$\tabularnewline
\hline 
$1000$ & $1.22\times10^{-2}$ & $1.22\times10^{-2}$ & $1.22\times10^{-2}$ & $1.19\times10^{-2}$\tabularnewline
\hline 
$pp\rightarrow bgX$ &  &  &  & \tabularnewline
\hline 
$500$ & $8.13\times10^{0}$ & $8.13\times10^{0}$ & $7.93\times10^{0}$ & $5.96\times10^{0}$\tabularnewline
\hline 
$600$ & $5.59\times10^{0}$ & $5.59\times10^{0}$ & $5.53\times10^{0}$ & $4.88\times10^{0}$\tabularnewline
\hline 
$700$ & $3.98\times10^{0}$ & $3.98\times10^{0}$ & $3.96\times10^{0}$ & $3.73\times10^{0}$\tabularnewline
\hline 
$800$ & $2.91\times10^{0}$ & $2.91\times10^{0}$ & $2.90\times10^{0}$ & $2.81\times10^{0}$\tabularnewline
\hline 
$900$ & $2.16\times10^{0}$ & $2.16\times10^{0}$ & $2.16\times10^{0}$ & $2.14\times10^{0}$\tabularnewline
\hline 
$1000$ & $1.64\times10^{0}$ & $1.63\times10^{0}$  & $1.63\times10^{0}$ & $1.62\times10^{0}$\tabularnewline
\hline 
$pp\rightarrow bZX$ &  &  &  & \tabularnewline
\hline 
$500$ & $7.87\times10^{-1}$ & $7.81\times10^{-1}$ & $7.50\times10^{-1}$ & $4.79\times10^{-1}$\tabularnewline
\hline 
$600$ & $5.48\times10^{-1}$ & $5.48\times10^{-1}$ & $5.31\times10^{-1}$ & $4.27\times10^{-1}$\tabularnewline
\hline 
$700$ & $3.95\times10^{-1}$ & $3.94\times10^{-1}$ & $3.86\times10^{-1}$ & $3.39\times10^{-1}$\tabularnewline
\hline 
$800$ & $2.92\times10^{-1}$ & $2.91\times10^{-1}$ & $2.86\times10^{-1}$ & $2.61\times10^{-1}$\tabularnewline
\hline 
$900$ & $2.18\times10^{-1}$ & $2.18\times10^{-1}$ & $2.15\times10^{-1}$ & $2.02\times10^{-1}$\tabularnewline
\hline 
$1000$ & $1.66\times10^{-1}$ & $1.66\times10^{-1}$ & $1.64\times10^{-1}$ & $1.56\times10^{-1}$ \tabularnewline
\hline 
\end{tabular}
\end{table}

\begin{table}
\protect\caption{The same as Table \ref{tab:tab11}, but for parametrization PII. \label{tab:tab12}}

\begin{tabular}{|c|c|c|c|c|}
\hline 
Signal & \multicolumn{4}{c|}{PII}\tabularnewline
\hline 
$pp\rightarrow b\gamma X$ & $p_{T}>20$ GeV & $p_{T}>50$ GeV & $p_{T}>100$ GeV & $p_{T}>200$ GeV\tabularnewline
\hline 
\hline 
$500$ & $5.18\times10^{-3}$ & $5.26\times10^{-3}$ & $5.04\times10^{-3}$ & $3.54\times10^{-3}$\tabularnewline
\hline 
$600$ & $3.38\times10^{-3}$ & $3.37\times10^{-3}$ & $3.36\times10^{-3}$ & $2.77\times10^{-3}$\tabularnewline
\hline 
$700$ & $2.32\times10^{-3}$ & $2.31\times10^{-3}$ & $2.30\times10^{-3}$ & $2.05\times10^{-3}$\tabularnewline
\hline 
$800$ & $1.71\times10^{-3}$ & $1.63\times10^{-3}$ & $1.64\times10^{-3}$ & $1.50\times10^{-3}$\tabularnewline
\hline 
$900$ & $1.17\times10^{-3}$ & $1.16\times10^{-3}$ & $1.17\times10^{-3}$ & $1.11\times10^{-3}$\tabularnewline
\hline 
$1000$ & $8.60\times10^{-4}$  & $8.58\times10^{-4}$  & $8.55\times10^{-4}$  & $8.24\times10^{-4}$ \tabularnewline
\hline 
$pp\rightarrow bgX$ &  &  &  & \tabularnewline
\hline 
$500$ & $7.40\times10^{-1}$ & $7.39\times10^{-1}$ & $7.21\times10^{-1}$ & $5.16\times10^{-1}$\tabularnewline
\hline 
$600$ & $4.83\times10^{-1}$ & $4.80\times10^{-1}$ & $4.81\times10^{-1}$ & $3.98\times10^{-1}$\tabularnewline
\hline 
$700$ & $3.22\times10^{-1}$ & $3.22\times10^{-1}$ & $3.20\times10^{-1}$! & $2.89\times10^{-1}$!\tabularnewline
\hline 
$800$ & $2.24\times10^{-1}$ & $2.21\times10^{-1}$ & $2.21\times10^{-1}$ & $2.04\times10^{-1}$\tabularnewline
\hline 
$900$ & $1.5\times10^{-1}$ & $1.58\times10^{-1}$ & $1.58\times10^{-1}$ & $1.49\times10^{-1}$\tabularnewline
\hline 
$1000$ & $1.14\times10^{-1}$  & $1.14\times10^{-1}$  & $1.13\times10^{-1}$  & $1.10\times10^{-1}$ \tabularnewline
\hline 
$pp\rightarrow bZX$ &  &  &  & \tabularnewline
\hline 
$500$ & $6.89\times10^{-2}$ & $6.85\times10^{-2}$ & $6.45\times10^{-2}$ & $4.23\times10^{-2}$\tabularnewline
\hline 
$600$ & $4.52\times10^{-2}$ & $4.51\times10^{-2}$ & $4.34\times10^{-2}$ & $3.53\times10^{-2}$\tabularnewline
\hline 
$700$ & $3.12\times10^{-2}$ & $3.11\times10^{-2}$ & $3.05\times10^{-2}$ & $2.65\times10^{-2}$\tabularnewline
\hline 
$800$ & $2.19\times10^{-2}$ & $2.18\times10^{-2}$ & $2.15\times10^{-2}$ & $1.95\times10^{-2}$\tabularnewline
\hline 
$900$ & $1.56\times10^{-2}$ & $1.56\times10^{-2}$ & $1.55\times10^{-2}$ & $1.44\times10^{-2}$\tabularnewline
\hline 
$1000$ & $1.14\times10^{-2}$  & $1.13\times10^{-2}$  & $1.13\times10^{-2}$  & $1.07\times10^{-2}$ \tabularnewline
\hline 
\end{tabular}
\end{table}

\begin{table}
\protect\caption{The same as Table \ref{tab:tab11}, but for parametrization PIII.
\label{tab:tab13}}

\begin{tabular}{|c|c|c|c|c|}
\hline 
Signal & \multicolumn{4}{c|}{PIII}\tabularnewline
\hline 
$pp\rightarrow b\gamma X$ & $p_{T}>20$ GeV & $p_{T}>50$ GeV & $p_{T}>100$ GeV & $p_{T}>200$ GeV\tabularnewline
\hline 
\hline 
$500$ & $13.1\times10^{-2}$ & $13.14\times10^{-2}$ & $12.75\times10^{-2}$ & $8.92\times10^{-2}$\tabularnewline
\hline 
$600$ & $8.59\times10^{-2}$ & $8.58\times10^{-2}$ & $8.44\times10^{-2}$ & $7.03\times10^{-2}$\tabularnewline
\hline 
$700$ & $5.82\times10^{-2}$ & $5.82\times10^{-2}$ & $5.77\times10^{-2}$ & $5.17\times10^{-2}$\tabularnewline
\hline 
$800$ & $4.07\times10^{-2}$ & $4.07\times10^{-2}$ & $4.06\times10^{-2}$ & $3.77\times10^{-2}$\tabularnewline
\hline 
$900$ & $2.92\times10^{-2}$ & $2.92\times10^{-2}$ & $2.91\times10^{-2}$ & $2.77\times10^{-2}$\tabularnewline
\hline 
$1000$ & $2.14\times10^{-2}$  & $2.13\times10^{-2}$  & $2.13\times10^{-2}$  & $2.06\times10^{-2}$ \tabularnewline
\hline 
$pp\rightarrow bgX$ &  &  &  & \tabularnewline
\hline 
$500$ & $19.04\times10^{0}$ & $18.96\times10^{0}$ & $18.43\times10^{0}$ & $12.86\times10^{0}$\tabularnewline
\hline 
$600$ & $12.19\times10^{0}$ & $12.13\times10^{0}$ & $11.93\times10^{0}$ & $9.92\times10^{0}$\tabularnewline
\hline 
$700$ & $8.08\times10^{0}$ & $8.07\times10^{0}$ & $8.02\times10^{0}$ & $7.17\times10^{0}$\tabularnewline
\hline 
$800$ & $5.57\times10^{0}$ & $5.57\times10^{0}$ & $5.55\times10^{0}$ & $5.15\times10^{0}$\tabularnewline
\hline 
$900$ & $3.94\times10^{0}$ & $3.94\times10^{0}$ & $3.94\times10^{0}$ & $3.74\times10^{0}$\tabularnewline
\hline 
$1000$ & $2.85\times10^{0}$  & $2.85\times10^{0}$  & $2.85\times10^{0}$  & $2.74\times10^{0}$ \tabularnewline
\hline 
$pp\rightarrow bZX$ &  &  &  & \tabularnewline
\hline 
$500$ & $1.76\times10^{0}$ & $1.75\times10^{0}$ & $1.65\times10^{0}$ & $1.05\times10^{0}$\tabularnewline
\hline 
$600$ & $1.15\times10^{0}$ & $1.14\times10^{0}$ & $1.11\times10^{0}$ & $8.80\times10^{-1}$\tabularnewline
\hline 
$700$ & $7.83\times10^{-1}$ & $7.80\times10^{-1}$ & $7.60\times10^{-1}$ & $6.61\times10^{-1}$\tabularnewline
\hline 
$800$ & $5.47\times10^{-1}$ & $5.41\times10^{-1}$ & $5.31\times10^{-1}$ & $4.80\times10^{-1}$\tabularnewline
\hline 
$900$ & $3.92\times10^{-1}$ & $3.90\times10^{-1}$ & $3.82\times10^{-1}$ & $3.60\times10^{-1}$\tabularnewline
\hline 
$1000$ & $2.86\times10^{-1}$  & $2.82\times10^{-1}$  & $2.80\times10^{-1}$  & $2.62\times10^{-1}$ \tabularnewline
\hline 
\end{tabular}
\end{table}

\begin{table}
\protect\caption{The cross sections (in pb) for the backgrounds ($b(\bar{b})V$, $c(\bar{c})V$
and $jV$, where $V=$photon, jet and $Z$ boson) with $p_{T}$ cuts
on the jets and photon at the center of mass energy $\sqrt{s}=13$
TeV. \label{tab:tab14}}

\begin{tabular}{|c|c|c|c|c|}
\hline 
Background & $p_{T}>20$ GeV & $p_{T}>50$ GeV & $p_{T}>100$ GeV & $p_{T}>200$ GeV\tabularnewline
\hline 
\hline 
$pp\to b(\bar{b})\gamma X$ & $2.99\times10^{3}$ & $1.35\times10^{2}$ & $9.04\times10^{0}$ & $4.02\times10^{-1}$\tabularnewline
\hline 
$pp\to c(\bar{c})\gamma X$ & $1.87\times10^{4}$ & $8.15\times10^{2}$ & $5.40\times10^{1}$ & $2.43\times10^{0}$\tabularnewline
\hline 
$pp\to j\gamma X$ & $5.43\times10^{4}$ & $3.27\times10^{3}$ & $3.38\times10^{2}$ & $2.85\times10^{1}$\tabularnewline
\hline 
$pp\to b(\bar{b})jX$ & $7.83\times10^{6}$ & $3.05\times10^{5}$ & $1.92\times10^{4}$ & $8.93\times10^{2}$\tabularnewline
\hline 
$pp\to c(\bar{c})jX$ & $1.22\times10^{7}$ & $4.55\times10^{5}$ & $2.89\times10^{4}$ & $1.35\times10^{3}$\tabularnewline
\hline 
$pp\to jjX$ & $2.43\times10^{8}$ & $8.54\times10^{6}$ & $5.44\times10^{5}$ & $2.80\times10^{4}$\tabularnewline
\hline 
$pp\to b(\bar{b})ZX$ & $5.02\times10^{2}$ & $1.35\times10^{2}$ & $2.25\times10^{1}$ & $1.56\times10^{0}$\tabularnewline
\hline 
$pp\to c(\bar{c})ZX$ & $5.96\times10^{2}$ & $1.58\times10^{2}$ & $2.64\times10^{1}$ & $1.83\times10^{0}$\tabularnewline
\hline 
$pp\to jZX$ & $8.00\times10^{3}$ & $2.08\times10^{3}$ & $4.08\times10^{2}$ & $4.12\times10^{1}$\tabularnewline
\hline 
\end{tabular}
\end{table}

\begin{figure}
\includegraphics{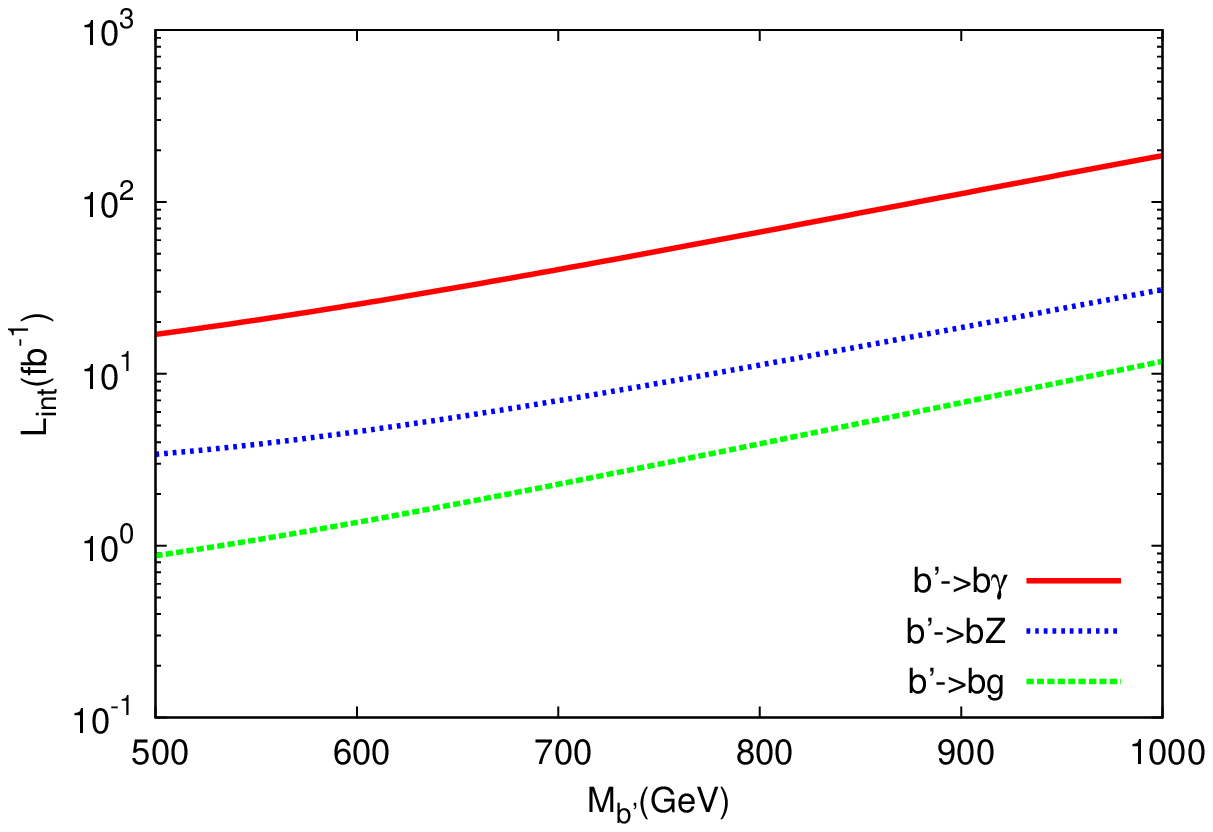}

\protect\caption{Integrated luminosity required to reach $3\sigma$ significance for
the signal of $b'$ anomalous interactions for parametrization PI
at the LHC with $\sqrt{s}=13$ TeV. \label{fig:fig8}}
\end{figure}

\begin{figure}
\includegraphics{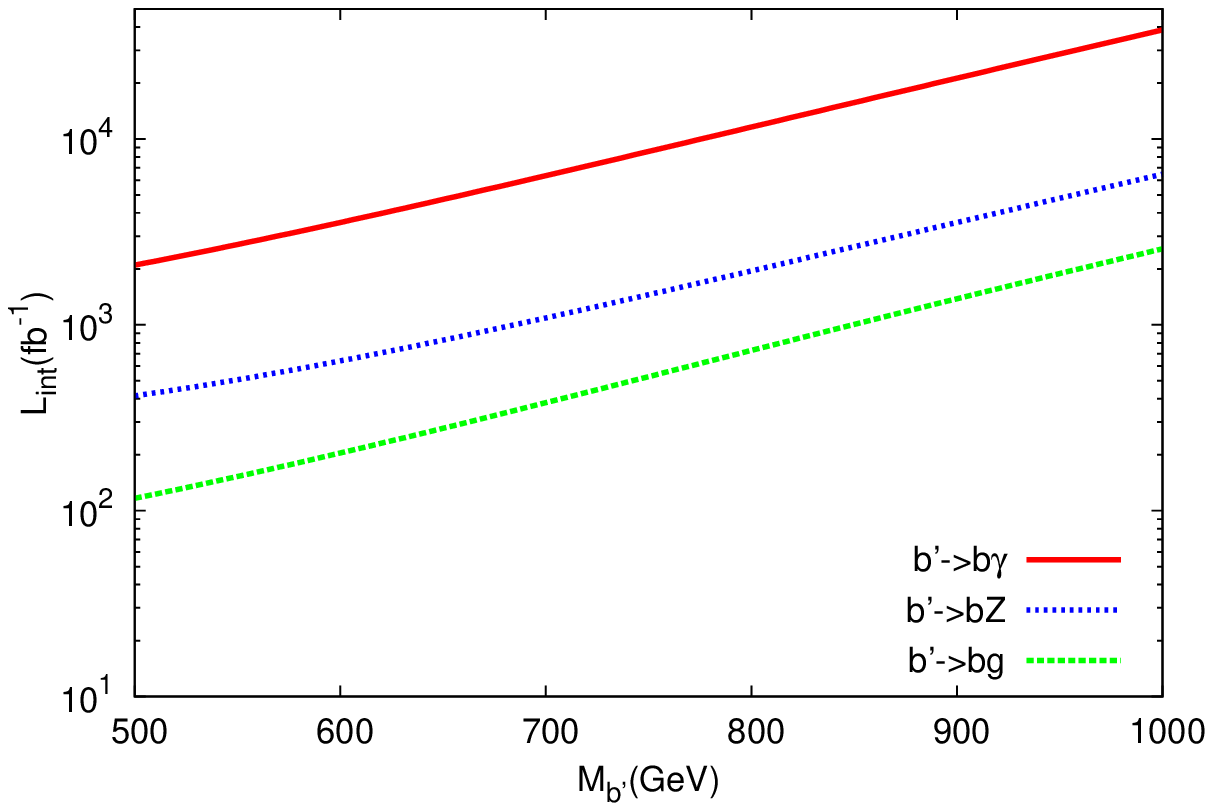}\protect\caption{The same as Fig. \ref{fig:fig8}, but for parametrization PII. \label{fig:fig9}}
\end{figure}

\begin{figure}
\includegraphics{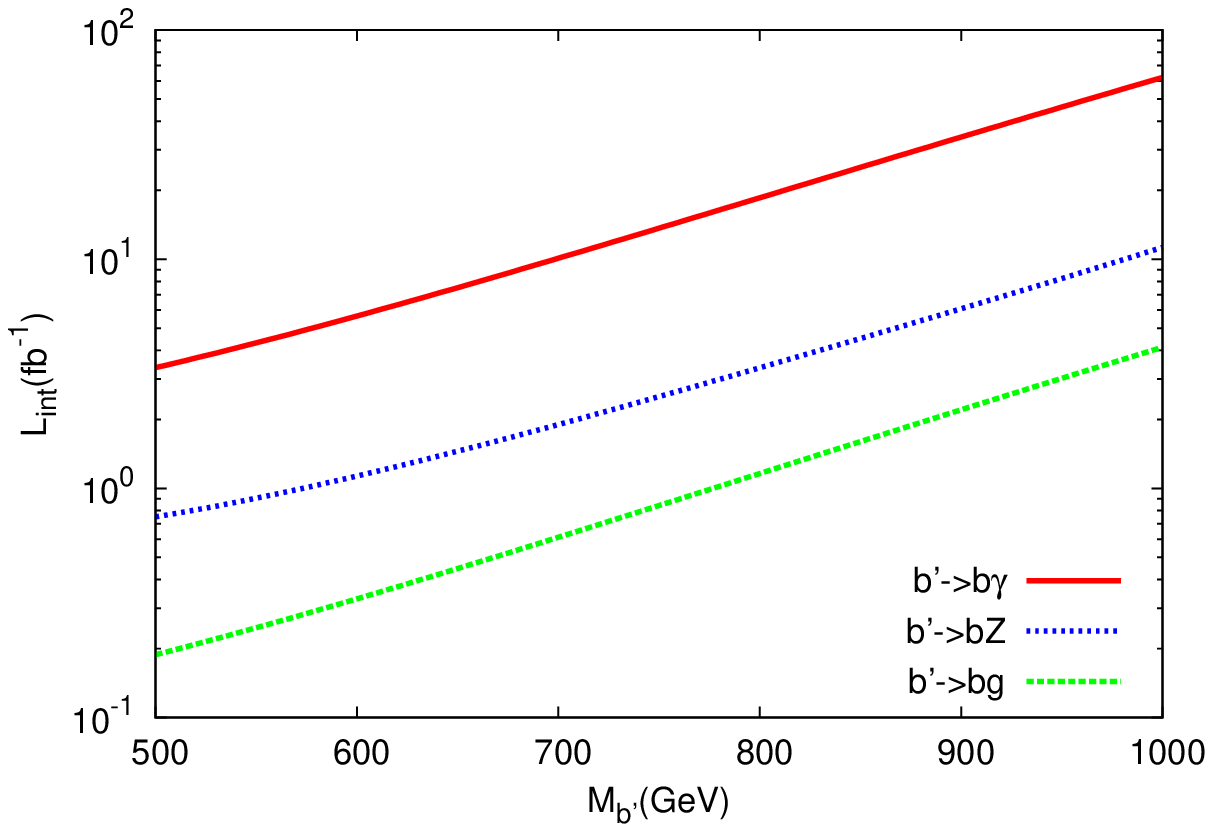}\protect\caption{The same as Fig. \ref{fig:fig8}, but for parametrization PIII. \label{fig:fig10}}
\end{figure}

\begin{figure}
\includegraphics[bb=0bp 0bp 250bp 244bp]{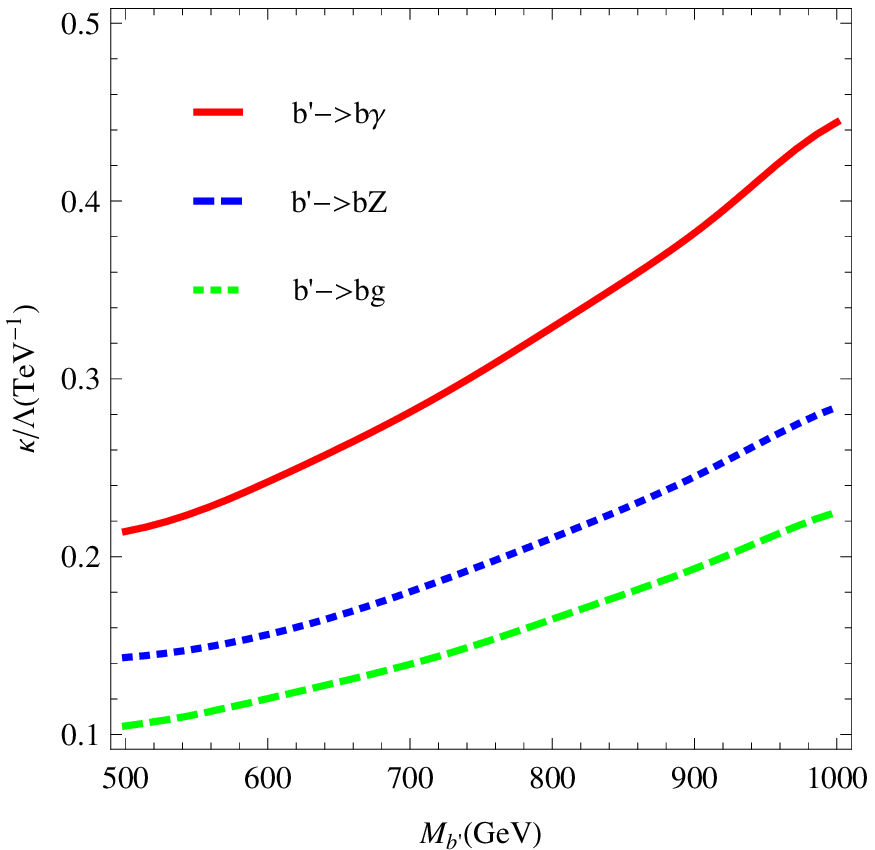}\protect\caption{The contour plot of anomalous coupling and mass of new heavy quark
$b'$ for the dynamical parametrization explained in the text with
a significance $3\sigma$ at $\sqrt{s}=13$ TeV and $L_{int}=100$
fb$^{-1}$. \label{fig:fig11}}
\end{figure}

In order to reach $3\sigma$ significance for the signal of $b'$
anomalous interactions the required integrated luminosity is shown
in Figs. \ref{fig:fig8}- \ref{fig:fig10} for parametrizations PI,
PII and PIII at the LHC with $\sqrt{s}=13$ TeV. The channel $b'\rightarrow b\gamma$
requires more integrated luminosity than the other channels. By requiring
the signal significance $SS=3$, the contour plots of $\kappa/\Lambda$
and mass of $b'$ quark are presented in Fig. \ref{fig:fig11}. The
results show that one can discover the $b'$ quark anomalous couplings
down to 0.1 in the $bg$ channel for $m_{b'}$=500 GeV. 

\section{Conclus\i on}

The new heavy quarks of up-type and down-type can be produced with
large numbers at the LHC if they have the anomalous couplings (via
flavour changing neutral current) that well dominate over the charged
current interactions. The single production of new heavy quarks can
be achieved through the anomalous interactions at the LHC with $\sqrt{s}=$13
TeV. The anomalous vertices could appear significantly at leading
order processes due to the possiblity of new heavy quarks. From the
results of signal significance calculations for $t'$ ($b'$) anomalous
productions, the sensitivity to the anomalous couplings $ $$\kappa^{t'}/\Lambda$
($\kappa^{b'}/\Lambda$) can be reached down to 0.1 TeV$^{-1}$ (0.15
TeV$^{-1}$) in the lepton+$b$-jet+jet+\emph{MET }($b$-jet+jet)
channel at $\sqrt{s}=$13 TeV, assuming a dynamical parametrization
for the anomalous couplings and the mass of 750 GeV for the new heavy
quarks. 
\begin{acknowledgments}
This work was supported in part by Turkish Atomic Energy Authority
(TAEA) under the project Grant No. 2011TAEKCERN-A5.H2.P1.01-19. \end{acknowledgments}

\end{document}